\documentclass[referee,sn-basic]{sn-jnl}
\newcommand{\norm}[1]{\left\lVert#1\right\rVert}
\usepackage{subcaption}
\usepackage{titleps}
\usepackage{url}
\usepackage{rotating}
\usepackage{graphicx}
\usepackage{booktabs}
\begin{document}

\title{Benchmarking distance-based partitioning methods for mixed-type data}


\author*[1]{\fnm{Efthymios} \sur{Costa}}\email{efthymios.costa17@imperial.ac.uk}

\author[1]{\fnm{Ioanna} \sur{Papatsouma}}\email{i.papatsouma@imperial.ac.uk}

\author[2]{\fnm{Angelos} \sur{Markos}}\email{amarkos@eled.duth.gr}

\affil*[1]{\orgdiv{Department of Mathematics}, \orgname{Imperial College London}} 

\affil[2]{\orgdiv{Department of Primary Education}, \orgname{Democritus University of Thrace}}


\abstract{Clustering mixed-type data, that is, observation by variable data that consist of both continuous and categorical variables poses novel challenges. Foremost among these challenges is the choice of the most appropriate clustering method for the data. This paper presents a benchmarking study comparing eight distance-based partitioning methods for mixed-type data in terms of cluster recovery performance. A series of simulations carried out by a full factorial design are presented that examined the effect of a variety of factors on cluster recovery. The amount of cluster overlap, the percentage of categorical variables in the data set, the number of clusters and the number of observations had the largest effects on cluster recovery and in most of the tested scenarios. KAMILA, K-Prototypes and sequential Factor Analysis and K-Means clustering typically performed better than other methods. The study can be a useful reference for practitioners in the choice of the most appropriate method.}


\maketitle

\section{Introduction}\label{sec1}

Benchmarking studies of clustering are increasingly important for guiding users and practitioners in choosing appropriate clustering approaches among an increasing number of alternatives \citep{whitepaper}. The objective of the present study is to contribute to the benchmarking literature by evaluating the performance of clustering methods for mixed-type data, that is, observation by variable data that consist of both continuous and categorical variables. Indeed, research in a variety of domains usually relies on heterogeneous or mixed-type data. In social science research, for example, data sets typically include demographic background characteristics (usually categorical variables) together with socioeconomic or psychological measures (usually continuous variables). Such heterogeneity urges for ways to guide users and practitioners in choosing appropriate clustering approaches for mixed-type data sets in order to identify distinct profiles of individuals and/or generate hypotheses, thereby contributing to a quantitative empirical methodology for the discipline.

Cluster analysis of mixed-type data sets can be a particularly challenging task because it requires to weigh and aggregate different variables against each other \citep{hennig2013find}. One of the main issues is the choice of the most appropriate distance or model to simultaneously process both data types. Among the most simple and intuitive strategies for clustering mixed-type data is to convert all variables to a single type, continuous or categorical, via discretization, dummy-coding or fuzzy-coding. Such a strategy may lead to a significant loss of information from the original data and may consequently lead to increased bias \citep{foss2018kamila}. Another approach is to cluster observations separately for continuous and categorical variables, and then match the clusters in the two clusterings. This approach, however, ignores any dependencies that might exist between variables of different types \citep{hunt2011clustering}. Fortunately, a wide range of clustering algorithms has been specifically developed to deal with mixed-type data. A taxonomy of available methods can be found in \cite{ahmad2019survey} and overviews of distance or dissimilarity-based methods are given by \cite{foss2019distance} and \cite{van2019distance}.



Benchmarking studies may be performed by independent groups interested in systematically comparing existing methods or by authors of new methods to demonstrate performance improvements or other advantages over existing competitors. With regard to studies performed by independent groups, there have been several benchmarking studies of clustering for continuous data only or categorical data only \citep[e.g.,][]{milligan1980,meil2001,ferreira2009,sarali2013,boulesteix2017,javed2020,hennig2022}, whereas benchmarking studies of clustering for mixed-type data are scarce  \citep{jimeno2021,preud2021}. We can also distinguish benchmarking studies of clustering for mixed-type data that are part of original papers where new methods are proposed  \citep[e.g.,][]{ahmad2007k,hennig2013find, foss2016semiparametric}. 

In this paper, we concern ourselves with distance or dissimilarity-based partitioning methods for mixed-type data, that is, methods that rely on explicit distances or dissimilarities between observations or between observations and cluster centroids. In the authors' understanding, these methods span three general approaches. The first approach involves computing an appropriate dissimilarity measure for mixed-type data, followed by a partitioning algorithm on the resulting dissimilarity matrix. Typically, such a dissimilarity measure can be constructed by defining and combining dissimilarity measures for each type of variable. A popular choice  in this category involves the computation of Gower's (dis)similarity measure \citep{gower1971general} among observations and then applying K-Medoids or hierarchical clustering on the dissimilarity matrix. Instead of using Gower's dissimilarity, \cite{hennig2013find} proposed a specific weighting scheme to more appropriately balance continuous against categorical variables. The second approach involves conducting a partitional clustering of the observation by variable data with the distances between observations and cluster centroids calculated separately for categorical and continuous variables, and combine them into a single objective function. Representative methods in this category are K-Prototypes \citep{huang1997clustering}, Modha-Spangler K-Means \citep{modha2003feature} and Mixed K-Means \citep{ahmad2007k}. The third approach comprises factor analysis of the variables and partitional clustering of the observations in the low-dimensional space. Factor analysis and clustering can be performed sequentially, i.e., in a two-step approach where clustering is applied to the resulting factor scores \cite[see e.g.,][]{dolnicar2008challenging} or simultaneously, where the two objectives are combined by optimizing a single convex objective function \citep{vichi2019clustering}. In the sequential approach, the first step usually involves Factor Analysis for Mixed Data or PCAMIX \citep{pages2014multiple, kiers1991simple} to obtain the observation scores in a low-dimensional space and then K-Means clustering to partition the observation scores. Simultaneous approaches include extensions of Reduced K-Means \citep{desoete1994k} and Factorial K-Means \citep{vichi2001factorial} to deal with the general relevant case of mixed variables \citep{vichi2019clustering}.

A simulation study was conducted to compare eight distance-based partitioning methods in terms of cluster recovery performance, following recommendations provided in \cite{boulesteix2013} and \cite{whitepaper}. The involved clustering methods represent a major class of methods listed in \cite{murtagh2015brief}, are well established and widely used - at least the less recent ones - while the most recently proposed have been shown in previous studies to outperform others. The study attempts to provide a neutral comparison of the methods since none of the authors have been involved in the development of any of the compared methods, and have no specific interest to portray any of them as particularly good or bad. The current study goes beyond previous work by considering different aspects that might affect performance (number of clusters, number of observations, number of variables, percentage of categorical variables in the data set, cluster overlap, cluster density and cluster sphericity), according to a full factorial design. The result is a concrete description of the method performance, with the goal of providing researchers with a guide to selecting the most suitable method for their study. 

The remainder of this paper is structured as follows: Section \ref{Sec:2} reviews benchmarking studies of clustering for mixed-type data, Section \ref{Sec:3} presents the methods under comparison, Section \ref{Sec:4} describes the simulation study design, Section \ref{Sec:5} presents the results and Section \ref{Sec:6} discusses the results and concludes the paper.

\section{Related work} \label{Sec:2}

Most comparison studies of clustering algorithms for mixed-type data have been performed within original articles presenting new methods, usually in order to establish their superiority over classical approaches. \cite{foss2016semiparametric} conducted a small scale simulation study and analyses of real-world data sets to illustrate the effectiveness of KAMILA, a newly proposed semi-parametric method, versus Modha–Spangler K-Means, K-Means with a weighting scheme described in \cite{hennig2013find} and two finite mixture models. They considered both normal and non-normal data sets (data following a $p$-generalized normal-multinomial distribution or the lognormal-multinomial distribution), with varying sample sizes (250, 500, 1000, and 10000), number of continuous variables (2, 4), number of categorical variables (1, 2, 3, and 4), level of continuous overlap (1\%, 15\%, 30\%, and 45\%), level of categorical overlap (1\%, 15\%, 30\%, 45\%, 60\%, 75\%, and 90\%), number of clusters (2, and 4), and number of categorical levels (2, and 4). The authors verified that the findings from the artificial data set analysis are generalizable in the context of real-world applications. KAMILA performed well across all conditions.

In line with \cite{foss2016semiparametric}, \cite{markos2020sequential} investigated the performance of three sequential dimensionality reduction and clustering approaches versus KAMILA, Modha-Spangler K-Means and Gower's (dis)similarity measure followed by Partitioning Around Medoids on simulated data with varying degree of cluster separation. More precisely, the study focused on three different scenarios; first, a scenario where both continuous and categorical variables have approximately comparable cluster overlap (i.e., contain equally useful information regarding the cluster structure), second, a scenario where continuous variables have substantially more overlap compared to categorical ones (i.e., the categorical variables are more useful for clustering purposes) and third, a case where categorical variables have substantially more overlap compared to continuous ones (i.e., the continuous variables are more useful for clustering purposes). They generated 500 data sets with 200 observations, two continuous and two categorical variables with four categories/levels each. Results showed that dimensionality reduction followed by clustering in the reduced space is an effective strategy for clustering mixed-type data when categorical variables are more informative than continuous ones with regard to the cluster structure.

More recently, \cite{jimeno2021} compared KAMILA, K-Prototypes and Multiple Correspondence Analysis (MCA) followed by K-Means, fuzzy C-Means, Probabilistic Distance Clustering or a mixture of Student's $t$ distributions, under different simulated scenarios. The study considered $27$ simulated scenarios based on three parameters: the number of clusters ($2$, $5$, and $7$), the amount of overlap in each cluster (30\%, 60\%, and 80\%), and the ratio of continuous variables to nominal variables (1:3, 1:1, and 3:1). The total numbers of variables and observations were fixed in each case ($128$ variables and $1920$ observations). K-Prototypes and KAMILA performed consistently well for spherical clusters. As the number of clusters increased, the performance of MCA followed by fuzzy C-means or PD clustering worsened.  MCA followed by a mixture of Student's $t$ distributions performed well in all cases.

A recent benchmarking study by \cite{preud2021} compared the performance of four model-based methods (KAMILA, Latent Class Analysis, Latent Class Model, and Clustering by Mixture Modeling) and five distance/dissimilarity-based methods (Gower's dissimilarity or Unsupervised Extra Trees dissimilarity followed by hierarchical clustering or Partitioning Around Medoids, K‐prototypes) on both simulated and real data. The parameters used for the simulations were the number of observations ($300$, $600$, and $1200$), the number of clusters ($2$, $6$, and $10$), the ratio between the number of continuous and categorical variables in the data, the proportion of relevant or non-noisy variables ($20$\%, $50$\%, and $90$\%), and the degree of relevance of the variables with regard to the cluster structure (low, mild, and high, as defined by a cluster separation index in the case of continuous variables and a noise proportion introduced via resampling in the categorical variables). The authors considered seven scenarios and $1000$ data sets were generated for each scenario. Results revealed the dominance of model-based over most distance or dissimilarity-based methods; this was somewhat expected since the simulated data matched the assumptions of model-based methods. K-Prototypes was the only efficient distance-based method, outperforming all other techniques for larger numbers of clusters. 

It is important to outline that none of the aforementioned studies made use of a full factorial design to enhance inferential capacity in terms of disentangling the effects of each of the manipulated parameters and their interactions. 


\section{Benchmark methods}\label{Sec:3}

The present study constrains its scope to distance or dissimilarity-based methods for partitioning mixed-type-data, i.e., methods that rely on explicit distances or dissimilarities between observations or between observations and cluster centroids. The methods under comparison produce crisp partitions, allow to fix the number of clusters in advance and have an R-implementation.

Two of the methods considered in the study involve the conversion of observation by variable data into observation by observation proximities. A popular choice is to calculate pairwise Gower's dissimilarities among observations \citep{gower1971general}:
\begin{equation}\label{eq:gowersdiss}
    d_{Gower}(\boldsymbol{X}_i, \boldsymbol{X}_{i'}) = 1- \frac{\sum\limits_{j=1}^p w_j(\boldsymbol{X}_i, \boldsymbol{X}_{i'})s_j(\boldsymbol{X}_i,\boldsymbol{X}_{i'})}{\sum\limits_{j=1}^p w_j(\boldsymbol{X}_i, \boldsymbol{X}_{i'})}, \quad 1 \leq i, i' \leq n,\ i\neq i',
\end{equation}
where $\boldsymbol{X}_i, \boldsymbol{X}_{i'}$ are distinct observations, therefore rows of an ($n \times p$)-dimensional data matrix. We denote the weight of the $j$th variable for the two observations by $w_j$; this is typically set to 1, assuming equal weight for all variables, but can also take different values for different variables based on their subject matter importance \citep{hennig2013find}. Finally, $s_j$ is a coefficient of similarity between the $j$th components of $\boldsymbol{X}_i$ and $\boldsymbol{X}_{i'}$, defined as the range-normalised Manhattan distance for continuous variables and the Kronecker delta for categorical ones. Gower's dissimilarity is very general and covers most applications of dissimilarity-based clustering to mixed-type variables.

In a discussion regarding the formal definition of `dissimilarity', \cite{hennig2013find} argue that a proper dissimilarity measure between data objects should not only aggregate the variables, but it should also make use of some weighting scheme that controls variable importance, especially for nominal variables. In fact, they propose standardising the continuous variables to unit variance, contrary to the range standardisation that is used for Gower's dissimilarity, while they introduce a more sophisticated weighting for nominal variables. More precisely, they claim that since it holds that for two independent and identically distributed continuous random variables $X_1$ and $X_2$, standardisation to unit variance implies $\mathrm{E}\left\{\left(X_1-X_2\right)^2\right\}=2$, standardisation of a nominal variable should be done in such a way that the dissimilarity between the two categories of a binary variable is about equal to the aforementioned expression or less than that, if the variable has more than two categorical levels. Based on this rationale, \cite{hennig2013find} suggest setting $\sum\limits_{i=1}^{I}\mathrm{E}\left\{\left(Z_{i1}-Z_{i2}\right)^2\right\}=\mathrm{E}\left\{\left(X_1-X_2\right)^2\right\}=2\xi$, where $Z_{i1},Z_{i2}$ represent the values of the first and second data points on the dummy variables $\boldsymbol{Z}_i$, obtained after dummy coding of a nominal variable with $I$ levels. The coefficient, $\xi$, is set to be equal to $1/2$ in order to avoid a clustering output that is highly dependent on the levels of the categorical variables. Thus, dummy-coded nominal variables are scaled so that they are comparable to unit variance scaled continuous variables. This is followed by constructing the Euclidean distance matrix between the observations, which is equivalent to the notion of a `dissimilarity matrix', as for Gower's dissimilarity.




Once the pairwise dissimilarities are calculated (either using Gower's or Hennig-Liao's measure), Partitioning Around Medoids or PAM \citep{kaufman2009finding} is applied to the proximity matrix obtained from the previous step. The objective of PAM is to find $K$ observations that will be representative, in the sense that they will minimise the average dissimilarity with all other points in each cluster. These are called `medoids' and are analogous to the `centroids' in the widely-used K-Means algorithm. In this study, the R package \texttt{cluster} \citep{Maechler2021} was used to calculate Gower's dissimilarity (function daisy()) and subsequently carry out PAM clustering (function pam()). The function distancefactor() of the R package \texttt{fpc} \citep{hennig2015packagefpc} was used for the standardisation of nominal variables based on \cite{hennig2013find}. Clustering mixed-type data with Gower's dissimilarity and the weighting scheme of \cite{hennig2013find} followed by PAM are herein referred to as `Gower/PAM' and `HL/PAM' respectively.

K-Prototypes is a clustering method introduced by \cite{huang1997clustering} for dealing with data of mixed type. The K-Means algorithm can be seen as a `special case' of this method, since the rationale behind it is that it seeks for a minimisation of the trace of the within cluster dispersion matrix cost function, defined as:
\begin{equation}\label{eq:Kprotcostfun}
    E = \sum\limits_{l=1}^K \sum\limits_{i=1}^n y_{il}\ d(\boldsymbol{X}_i, \boldsymbol{Q}_l).
\end{equation}
The term $y_{il}$ in Equation \eqref{eq:Kprotcostfun} denotes the $(i,l)$th element of an $(n \times K)$ partition matrix, taking values 0 and 1 (1 indicating cluster membership), while $\boldsymbol{Q}_l$ is the prototype for the $l$th cluster. A prototype is the equivalent to a medoid for PAM or a centroid for K-Means. The distance between $\boldsymbol{X}_i$ and $\boldsymbol{Q}_l$ is denoted by $d(\boldsymbol{X}_i, \boldsymbol{Q}_l)$ and it is calculated as a combination of the squared Euclidean distance for continuous and the weighted binary indicator for categorical variables. Assuming, without loss of generality, that the first $p_r<p$ variables in our data set are continuous and the rest are categorical, this may be expressed as:
\begin{equation}\label{eq:mixeddistanceKprot}
    d(\boldsymbol{X}_i, \boldsymbol{Q}_l)= \sum\limits_{j=1}^{p_r}(x_{ij} - q_{lj})^2 + \gamma_l\sum\limits_{j=p_r+1}^{p}\delta (x_{ij}, q_{lj}).
\end{equation}
In the expression above, $\gamma_l$ is a weight coefficient for categorical variables in the $l$th cluster; setting it to zero (thus indicating the absence of categorical variables), one can recover the K-Means algorithm. For computational reasons, the value of $\gamma_l$ is chosen to be the same for all clusters and it is calculated as the ratio of the variance of continuous variables to the variance of the categorical variables in the data set. For the $j$th categorical variable, the variance is defined as $1-\sum\limits_h p_{jh}^2$, where $p_{jh}$ is the frequency of $h$th categorical level of the $j$th variable divided by $n$. It can be shown that the components $q_{lj}$ ($j=1, \dots, p)$ of $\boldsymbol{Q}_l$ upon minimisation of \eqref{eq:Kprotcostfun} are given by the mean or the mode of values that the $j$th variable takes in the $l$th cluster, for $j$ being a continuous or a categorical variable respectively. K-Prototypes was conducted in this study using the kproto() function in the R package \texttt{clustMixType} \citep{Szepannek2018}.

Some of the shortcomings of K-Prototypes, as argued by \cite{ahmad2007k}, include the use of the mode of categorical variables while ignoring other frequent categories, the fact that the distance for categorical variables is not weighted and the need for a more `refined' notion of categorical distance. Therefore, \cite{ahmad2007k} proposed another K-Means-based algorithm for mixed-type data, that scales the Euclidean distance and calculates categorical distances based on the co-occurrence of categorical values (herein referred to as `Mixed K-Means').

More precisely, the distance between two distinct categories is first calculated with respect to the rest of the variables. Say $A$ and $B$ are two categories of the same $j$th categorical variable, then the categorical distance between the two with respect a $j'$th categorical variable is defined as the sum of the conditional probability that the $j$th component of an observation $\boldsymbol{X}_i$ takes the value $A$, given that $x_{ij'}$ is in some subset $\sigma$ of possible values of a $j'$th variable and the conditional probability that $x_{ij}=B$ given $x_{ij'} \notin \sigma$. In order for this to be a distance metric satisfying that the distance between two identical categorical values is zero, we subtract a unit from the sum we obtain. Notice that $\sigma$ is chosen carefully among all possible subsets of values of a $j'$th variable so that it maximises the aforementioned sum. Then, the distance between $A$ and $B$ is calculated as the average of all the categorical distances with respect to all other variables. Since this notion is defined for a $j'$th categorical variable, \cite{ahmad2007k} suggested a simple algorithm for the discretization of continuous variables in intervals of equal width. This discretization is also used for determining the weight of continuous variables, which is defined to be the average categorical distance between all possible combinations of the categorical levels introduced. 

To conduct Mixed K-Means in this study, a distance matrix was first computed using the function distmix() from the R package \texttt{kmed} and was then supplied as the input
to a K-Medoids algorithm, implemented in the function fastkmed() of the same package.

Another clustering method, that is based on K-Prototypes, is Modha-Spangler K-Means \citep{modha2003feature}. Once again, the Euclidean distance is used for continuous variables (not scaled, unlike for Mixed K-Means), while the cosine dissimilarity is used for categorical variables. The objective function is thus given by:
\begin{equation}\label{eq:modhaspanglerdist}
    d_{MS}(\boldsymbol{X}_i,\boldsymbol{Q}_l)=\sum\limits_{j=1}^{p_r}(x_{ij}-q_{lj})^2 + \gamma_l \left( 1-\frac{\sum\limits_{j=p_r+1}^{p^*} x_{ij}q_{lj}}{\sqrt{\sum\limits_{j=p_r+1}^{p^*} x_{ij}^2}\sqrt{\sum\limits_{j=p_r+1}^{p^*} q_{lj}^2}}\right)
\end{equation}
which is really similar to the cost function \eqref{eq:Kprotcostfun} but uses a different categorical distance. For the cosine dissimilarity to be used as in Equation \eqref{eq:modhaspanglerdist}, we need to make sure that our categorical variables are first dummy-coded, so that inner products and norms of vectors can be calculated. We also denote the total number of columns for continuous and dummy-coded categorical variables by $p^*$.

Modha-Spangler K-Means is a convex algorithm as both distance functions used are convex. One of its main strengths is that the coefficient $\gamma_l$ is automatically determined by the algorithm, by trying to minimise the ratio of the product of the average within-cluster dispersion for continuous and categorical variables to the product of the average between-cluster dispersion for continuous and categorical variables. A much more detailed description can be found in \cite{modha2003feature}. However, a weakness of this algorithm is that it requires a brute-force approach for determining the optimal value of $\gamma_l$; usually a greedy search over a grid of values specified by the user is employed. In our case, due to computational constraints, we consider only five candidate values of $\gamma_l$, which are the values of the set $\Gamma_l = \{ \frac{i}{6} : i \in [1, 5] \}$. The element of $\Gamma_l$ that yields the smallest value for the objective function \eqref{eq:modhaspanglerdist} is the one that is eventually used for $\gamma_l$. Modha-Spangler K-Means was applied in this study using the function gmsClust() of the R package \texttt{kamila} \citep{foss2018kamila}. The number of distinct cluster weightings evaluated in the brute-force search was set to $10$ (the default option).

The five aforementioned clustering methods all work with the full data, in perhaps very high dimensions. Another approach to cluster analysis is the so-called `tandem analysis', a term coined by \cite{arabie1994cluster}, which consists of a dimensionality reduction step via factor analysis, followed by a clustering of the observations in the resulting low-dimensional space \citep[see also][]{dolnicar2008challenging}. One such dimensionality reduction method, suitable for mixed-type data, is Factor Analysis for Mixed Data or FAMD (also known as Principal Component Analysis for Mixed Data) \citep{pages2014multiple}. 

Dimensionality reduction in FAMD is seen as a compromise between Principal Component Analysis and Multiple Correspondence Analysis \citep{markos2020sequential}. The idea is that the data matrix is partitioned in such a way that all columns consisting of continuous variables are `stacked' right next to an `indicator matrix' or `complete disjunctive table' that is constructed from the categorical variables. This partition matrix is constructed by recoding the categorical variables using dummy variables. The usual standardisation process of subtracting from each column its mean and dividing by its standard deviation is used for continuous variables. Standardisation of the indicator matrix is achieved by dividing the elements of each of its columns by the square root of the proportion of observations possessing the respective category that the column represents. Then, the two standardised matrices are concatenated and standard Principal Component Analysis (PCA) is performed on the resulting matrix. As in PCA, when applying FAMD it is important to decide on the number of factors to retain.

\cite{audigier2016principal} state that if the $i$th principal component obtained is denoted by $\boldsymbol{F}_i$, then the first principal component $\boldsymbol{F}_1$ maximises the expression:
\begin{equation}\label{eq:famdpca1max}
    \sum\limits_{j=1}^{p_r}R^2\left(\boldsymbol{F}_i, \boldsymbol{X}_{con_j} \right) + \sum\limits_{j=p_r+1}^p \eta^2\left(\boldsymbol{F}_i, \boldsymbol{X}_{cat_j} \right).
\end{equation}
Here, $R^2$ represents the coefficient of determination and $\eta^2$ is the squared correlation ratio, also known as the `Intraclass Correlation Coefficient'. Moreover, $\boldsymbol{X}_{con_j}$ and $\boldsymbol{X}_{cat_j}$ denote the $j$th continuous and categorical variables respectively, with $j$ being the column index. Maximising Equation \eqref{eq:famdpca1max} is therefore equivalent to maximising the link between continuous and categorical variables, so one may view $\boldsymbol{F}_1$ as the synthetic variable that is most correlated with both continuous and categorical variables. Similarly, $\boldsymbol{F}_2$ will be the synthetic variable orthogonal to $\boldsymbol{F}_1$ maximising Equation \eqref{eq:famdpca1max}. Once the dimensionality reduction step has been implemented, the final step consists of applying K-Means clustering on the lower-dimensional representation that has been obtained. This two-step procedure is herein referred to as `FAMD/K-Means'. FAMD is conducted using the function FAMD() of the R package \texttt{FactoMineR} and K-Means using the base R function kmeans().

While FAMD followed by K-Means and generally tandem analysis seems like a reasonable approach to the clustering problem, \cite{desoete1994k} raise the point that variables with little contribution to the cluster structure can potentially `mask' this structure, thus leading to unreliable results. This problem of `cluster masking' is described in more detail by \cite{vichi2019clustering}, who provide an illustration via a toy example. The idea of performing dimensionality reduction via PCA and K-Means clustering simultaneously, known as Reduced K-Means, was introduced in \cite{desoete1994k} as a potential solution to the cluster masking problem, with \cite{van2017cluster} giving a concise description of a similar algorithm suitable for categorical data. \cite{vichi2019clustering} generalized Reduced K-Means algorithm in the case of mixed-type data.

This joint dimensionality reduction and clustering technique is referred to as Mixed Reduced K-Means, where `Mixed' indicates the presence of mixed-type data \citep{van2019distance}. Its objective function is given by:
\begin{equation}\label{eq:rkmobjectivefun}
    \phi_{RKM}\left( \boldsymbol{B}, \boldsymbol{Z}_K, \boldsymbol{G}\right) = \norm{\boldsymbol{X}-\boldsymbol{Z}_K\boldsymbol{G}\boldsymbol{B}^\intercal}_F^2,
\end{equation}
with $\boldsymbol{X}, \boldsymbol{Z}_K, \boldsymbol{G}, \boldsymbol{B}$ indicating the data matrix is centered and standardised in the exact same way as described for FAMD, the $(n \times K)$-dimensional cluster membership matrix, the $(K \times d)$-dimensional matrix of cluster centroids in the reduced $d$-dimensional space and a $\left(p^* \times d\right)$-dimensional columnwise orthonormal loadings matrix respectively. We use $\norm{\cdot}_F$ to refer to the Frobenius norm and Equation \eqref{eq:rkmobjectivefun} is minimised via an Alternating Least Squares algorithm. In fact, it can be shown that there exists a certain expression for $\boldsymbol{G}$ that minimises \eqref{eq:rkmobjectivefun}, from which one can derive an expression for $\phi_{RKM}$ that only depends on $\boldsymbol{Z}_K$ and $\boldsymbol{B}$. The ALS algorithm used will first update the loadings matrix $\boldsymbol{B}$ while keeping $\boldsymbol{Z}_K$ fixed and this corresponds to a dimensionality reduction step. Once $\boldsymbol{B}$ has been updated, it is kept fixed and $\boldsymbol{Z}_K$ is updated accordingly, which can be seen as a K-Means problem. This also explains the intuition behind this algorithm performing joint dimensionality reduction and clustering. The choice of the number of dimensions retained, namely $d$, is set to be equal to $K-1$, where $K$ is the number of clusters. This follows from the recommendation of \cite{vichi2001factorial}, who argue that keeping more than $K-1$ dimensions is wasteful in joint dimensionality reduction and clustering algorithms, as this corresponds to describing a low-dimensional configuration of centroids in more dimensions than necessary. We have used the same number of dimensions in FAMD as in Mixed RKM for consistency. Notice that while Mixed RKM, as well as Mixed Factorial K-Means, which we have not implemented in our study, seem like reasonable methods for one to implement, \cite{yamamoto2014general} warn that these joint dimensionality reduction and clustering techniques are prone to giving inaccurate results if there exist variables irrelevant to the cluster structure in the data, which also happen to have high correlations between each other. Reduced K-Means was conducted in the current study using the function cluspca() of the R package \texttt{clustrd} \citep{markos2019beyond}.

The final method considered in the study is the KAMILA (KAy-means for MIxed LArge data) algorithm, that was introduced in \cite{foss2016semiparametric}. The method attempts to cluster mixed-type data while balancing the level of contribution of continuous and categorical variables in a flexible way, such that no strong parametric assumptions are made. KAMILA can be seen as a combination of the K-Means algorithm with the Gaussian-Multinomial mixture model that is commonly used in model-based clustering \citep{hunt2011clustering}. In fact, the crucial assumption of Gaussianity of the continuous variables $\boldsymbol{X}_{con}=\left(\boldsymbol{X}_{{con}_1}, \dots, \boldsymbol{X}_{{con}_{p_r}}\right)^\intercal$ is relaxed in KAMILA by considering the following more general probability density function for spherically symmetric distributions centered at the origin:
\begin{equation}\label{eq:kamilapdf}
    f_{\boldsymbol{X}_{con}}\left(\boldsymbol{x}_{con}\right)= \frac{f_R(r)\Gamma\left(\frac{p_r}{2}+1\right)}{p_rr^{p_r-1}\pi^{\frac{p_r}{2}}}.
\end{equation}
Equation \eqref{eq:kamilapdf} requires the evaluation of the density of pairwise distances between continuous variables $r=\sqrt{\boldsymbol{x}_{con}^\intercal \boldsymbol{x}_{con}}$, which is replaced by a univariate kernel density estimate that uses the Radial Basis Function (RBF) kernel. The assignment of data points in the clusters is made according to the sum of the estimated $\hat{f}_{\boldsymbol{X}_{con}}$ log likelihood value at the Euclidean distance of each data point to each cluster centroid and of the log probability of observing the $i$th categorical vector given population membership. This quantity needs to be calculated for each cluster separately, with the cluster maximising the quantity being the one that the data point will be assigned to. Notice that this formulation is valid under the assumption of independence of the $p-p_r$ categorical variables within each cluster. The iterative scheme of KAMILA updates the cluster centroids and the parameters of the assumed underlying multinomial and spherically symmetric distributions until these remain unchanged, thus yielding the same partitions. Although KAMILA cannot be strictly considered as a distance-based partitioning approach, it was included in the comparison as a model-based alternative of K-Means. Moreover, in a series of studies it was consistently found to outperform model-based and distance-based methods. A more detailed mathematical description of the algorithm can be found in \cite{foss2016semiparametric}, while the R implementation of KAMILA is available in the \texttt{kamila} package (function kamila()) \citep{foss2018kamila}.

\section{Simulation Study}\label{Sec:4}

A simulation study was conducted to evaluate the performance of the  eight clustering methods presented in the previous Section in terms of cluster recovery. The data were generated using the function MixSim() of the R package \texttt{MixSim} \citep{melnykov2012}, which allows for the determination of pairwise overlap between any pair of clusters. The notion of pairwise overlap is defined as the sum of two misclassification probabilities for a pair of weighted Gaussian distributions \citep{melnykov2010}. More precisely, if $\boldsymbol{X}$ is a random variable originating from cluster $l'$, the probability that it is misclassified to be originating from the $l$th cluster is given by $\omega_{l \lvert l'} = \mathbb{P}_{\boldsymbol{X}}\left( \pi_{l'}\phi\left(\boldsymbol{X}; \boldsymbol{\mu}_{l'}, \boldsymbol{\Sigma}_{l'}\right) < \pi_l\phi\left(\boldsymbol{X}; \boldsymbol{\mu}_l, \boldsymbol{\Sigma}_l\right) \lvert \boldsymbol{X} \sim \mathcal{N}_p\left(\boldsymbol{\mu}_{l'}, \boldsymbol{\Sigma}_{l'} \right) \right)$. Defining $\omega_{l' \lvert l}$ analogously, the overlap between the two clusters is given by $\omega_{ll'} = \omega_{l \lvert l'} + \omega_{l' \lvert l}$. Notice that $\boldsymbol{\mu}_l, \boldsymbol{\mu}_{l'}, \boldsymbol{\Sigma}_l$ and $\boldsymbol{\Sigma}_{l'}$ denote the mean vectors and the covariance matrices of the $l$th and the $l'$th components. However, this notion is only determined for continuous variables. In order to generate a categorical variable, a continuous variable was discretized by dividing it into $c$ classes with the 100/c\% quantile as the cut point. For simplicity, we will be assuming an equal number of categorical levels for all categorical variables (set equal to $4$).

Seven factors that are typical of data sets collected in real-world scenarios and commonly encountered in benchmarking studies of clustering, were systematically manipulated for data generation. The first factor, the number of clusters in the data set, was examined at three levels, $K = 3, 5,$ and $8$. The second factor, the number of observations, was evaluated at three levels, $n = 100, 600$ and $1000$, corresponding to a small, moderately large and large sample size in the social and behavioral sciences. The third factor, number of variables, was tested at three levels, $p = 8, 12$ and $16$. The fourth factor, overlap of clusters, assumed values of $0.1$\%, $0.5$\%, $1.0$\%, $1.5$\% and $2.0$\%, corresponding to very small, small, moderate, high and very high overlap (specified via the argument \texttt{BarOmega} of the function MixSim()). These values correspond to smaller overlap, similar overlap, and much more overlap than in real data sets typically used to demonstrate clustering algorithms (see \cite{shireman2016local}, for a justification and discussion of cluster overlap in real data sets compared to simulated data sets). The fifth factor, percentage of categorical variables in the data (versus continuous variables), was tested at three levels, $20$\%, $50$\% and $80$\%. The sixth factor, density of the clusters, was tested at two levels: (a) an equal number of observations in each cluster and (b) $10$\% of the observations in one cluster and the remaining observations equally divided across the remaining clusters. The seventh factor considered is cluster sphericity, defined by the covariance matrix structure with two levels: (a) mixtures of heteroscedastic spherical components, that is spherical covariance matrices with nonhomogeneous variances among the mixture components (arguments \texttt{sph}  and \texttt{hom} of the function MixSim() were set to \texttt{TRUE} and \texttt{FALSE}, respectively) and (b) mixtures of heteroscedastic non-spherical components, that is ellipsoidal covariance matrices with nonhomogeneous variances among the mixture components (arguments \texttt{sph} and \texttt{hom} were both set to \texttt{FALSE}). This resulted in $3 \times 3 \times 3 \times 5 \times 3 \times 2 \times 2 = 1620$ distinct data scenarios. Fifty replications were made for each scenario, resulting in a total of $81000$ data sets. Each clustering procedure was fit $100$ times using random starting values. For each data set, the number of clusters was always correctly specified. For FAMD and Mixed Reduced K-Means the number of dimensions (factors) was set to the number of clusters minus one. The ability of each procedure to return the true cluster structure was measured by the Adjusted Rand Index \citep[ARI;][]{hubert1985} and the Adjusted Mutual Information \citep[AMI;][]{vinh2010information}. The ARI measures the agreement between two different partitions of the same set of observations, by looking at pairs of observations in the original data set and counting and comparing how many pairs were assigned to the same cluster in both partitions, and how many pairs were not assigned to the same clusters in both partitions. The maximum value of the ARI is 1 and its expected value in the case of random partitions is 0. \cite{steinley2004properties} has provided some guidelines for interpreting ARI values in simulation experiments, with thresholds of $.90$, $.80$, and $.65$ corresponding to excellent, good, and fair cluster recovery, respectively. Values of the ARI below $.65$ reflect poor recovery. The AMI is an information-theoretic index that measures the amount of ``shared information" between two clusterings and is expected to be less susceptible to cluster size imbalance than ARI \citep{vanderhoef2019}. Both ARI and AMI measure the similarity between ground truth class assignments and those of the clustering method, adjusted for chance groupings. The simulated data sets, the resulting ARI and AMI values and the R code used for analyses are publicly available in an OSF repository at \url{https://rb.gy/rgpdyu}.

\section{Results}\label{Sec:5}

Table \ref{Tab:1} reports average cluster recovery of the eight methods across all factors. The best performing methods, on average, are KAMILA, FAMD/K-Means and K-Prototypes, followed by Modha-Spangler K-Means and Mixed Reduced K-Means. The worst performing methods were HL/PAM, Mixed K-Means and Gower/PAM. Table \ref{Tab:1} also presents the degree of agreement in cluster recovery between methods in terms of ARI/AMI, based on Pearson's correlation. Some pairwise correlations are large enough, the largest being Cor(FAMD/K-Means, Mixed Reduced K-Means) = $.94$/$.96$; that is both FAMD/K-Means and Mixed Reduced K-Means account for $88$\%/$92$\% of the variance in the other.


\begin{table}[h!]
\begin{center}
\begin{minipage}{300pt}
\caption{Agreement between methods based on Pearson's correlation and mean cluster recovery (ARI/AMI values) in the analysis of simulated data sets.}\label{Tab:1}%
\centering
\rotatebox{90}{
\begin{tabular}{@{}lcccccccc@{}}
 \toprule
 Method & (1) & (2) & (3) & (4) & (5) & (6) & (7) &  Mean\\ARI/AMI \\
  \midrule
 (1) KAMILA & - &  &  &  &  & & & .366/.404 \\[0.3cm]
 (2) FAMD/K-Means  & .90/.93 & - & & & & & & .340/.381 \\[0.3cm]
 (3) K-Prototypes & .93/.95  & .90/.92  & -& & & & & .336/.377 \\[0.3cm]
 (4) M-S K-Means & .90/.91 & .89/.91 & .88/.91 & -& & & & .313/.357 \\[0.3cm]
 (5) Mixed RKM & .87/.90 & .94/.96 & .88/.89 & .92/.93 & -& & & .309/.357  \\[0.3cm]
 (6) HL/PAM & .79/.82 & .75/.78 & .81/.86 & .78/.82 & .76/.79 & -& &  .193/.246 \\[0.3cm]
 (7) Mixed K-Means & .78/.82 & .70/.74 & .81/.86 & .75/.80 & .69/.74 & .81/.88 & -&  .191/.246 \\[0.3cm]
 (8) Gower/PAM & .73/.72 & .75/.75 & .73/.74 & .74/.75 & .77/.77 & .65/.71 & .77/.82 & .136/.182 \\[0.3cm]
\botrule
\end{tabular}}
\end{minipage}
\end{center}
\end{table}
\newpage

\begin{figure}[H]
  \centering
  \includegraphics[scale=0.4]{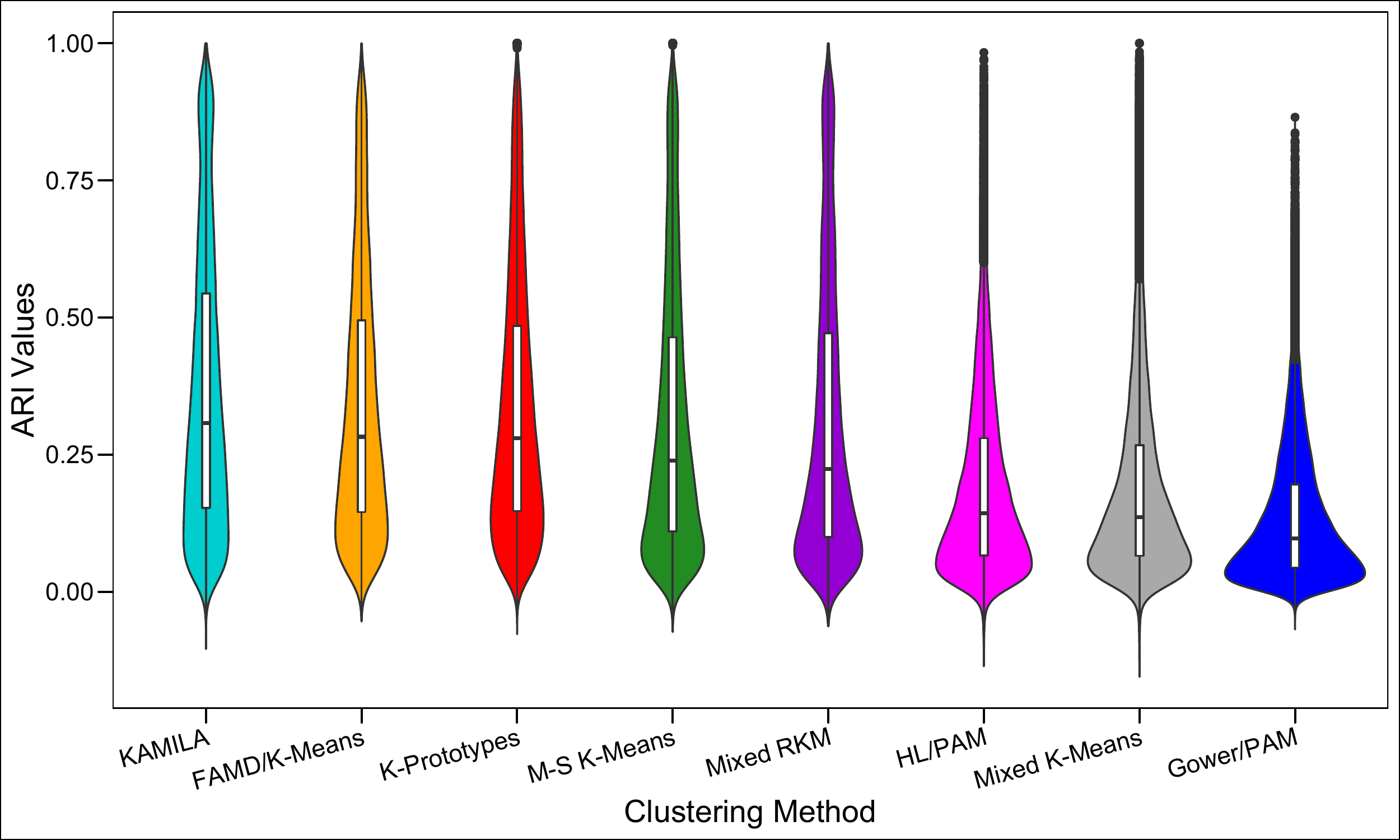}
  \caption{Violin/box plots of Adjusted Rand Index values by method}

  \label{Fig:1}
\end{figure}

\begin{figure}[H]
  \centering
  \includegraphics[scale=0.4]{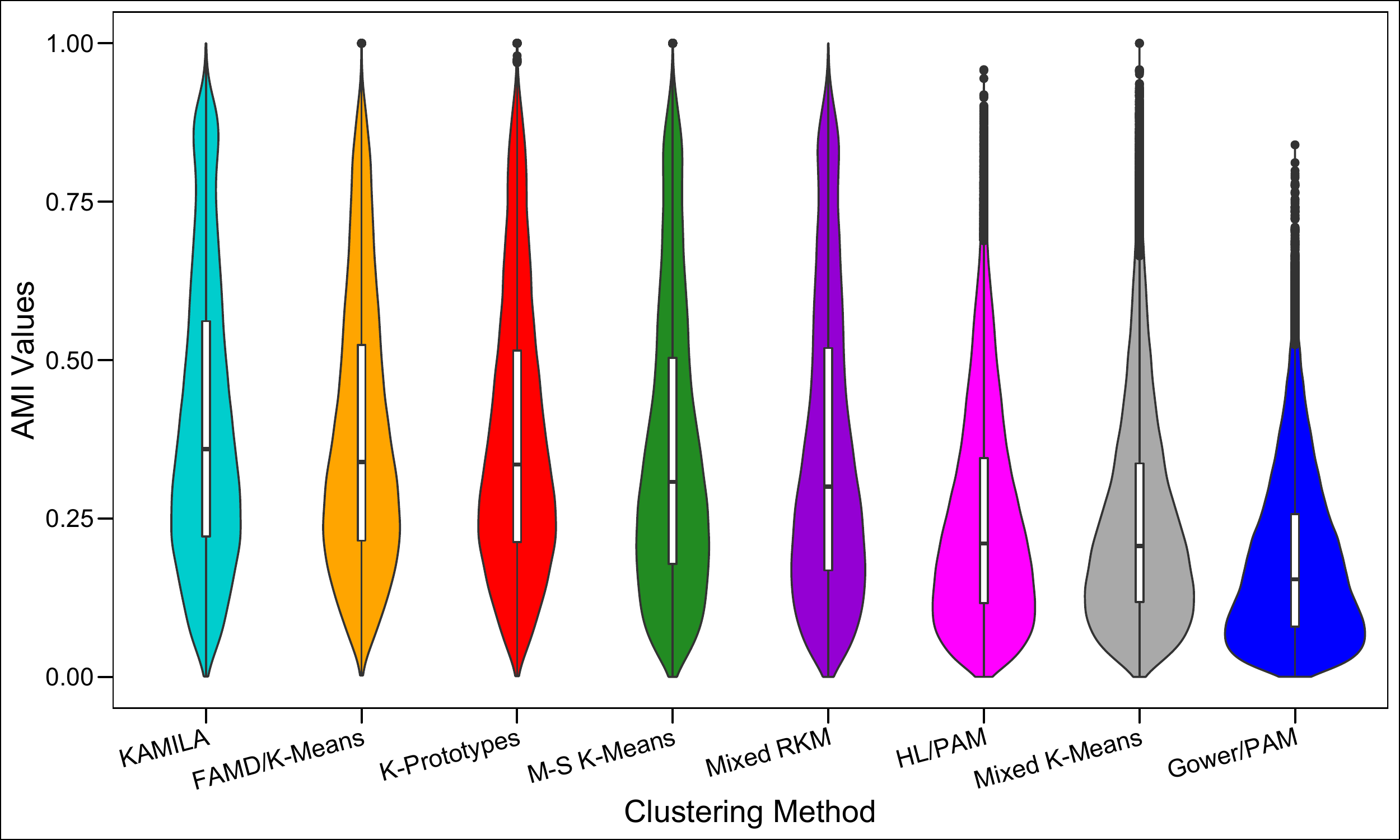}
  \caption{Violin/box plots of Adjusted Mutual Information values by method}

  \label{Fig:2}
\end{figure}

The violin/box plots in Figures \ref{Fig:1} and \ref{Fig:2} show the corresponding distributions of ARI and AMI values, respectively, for the eight methods computed on the true clusterings, confirming that KAMILA, FAMD/K-Means and K-Prototypes  perform somewhat better than Modha-Spangler K-Means and Mixed Reduced K-Means, and much better than HL/PAM, Mixed K-Means and Gower/PAM. For instance, more than $50$\% of the ARI values for HL/PAM, Mixed K-Means and Gower/PAM are less than $.14$. Also notice that more than $75$\% of the ARI values for Gower/PAM are less than $.20$, indicating poor cluster recovery.

After examining the overall performance of the methods, it is informative to determine if method performance is dependent upon specific situations (i.e., performance varies with the factor levels). The individual performances are examined by the levels of each factor. For this purpose, two separate repeated-measures ANOVAs were conducted on ARI and AMI scores (see Table \ref{Tab:2}). All main effects and interactions were modeled. Given the large sample size, it was expected that most factors would be statistically significant; therefore, all effects were evaluated with respect to their estimated effect sizes, partial eta-squared ($\eta^2$). Main effects and interactions were presented and discussed further only if they reached at least a moderate effect size (partial $\eta^2 \geq .01$).

\begin{table}[]
\caption{Repeated measures ANOVAs for eight clustering methods on ARI (top half) and AMI (bottom half). Factors are ordered by decreasing effect size, partial $\eta^2$.}
\begin{center}
\begin{minipage}{400pt}
\begin{tabular}{lllrrrr}
 \toprule
& Effect & Source & df      & SS      & $F$        & Partial $\eta^2$ \\ \midrule
\multirow{15}{*}{\begin{tabular}[c]{@{}l@{}}ARI\end{tabular}} & \multirow{7}{*}{\begin{tabular}[c]{@{}l@{}}Between\\ data sets\\ effects\end{tabular}}                       & overlap     & 4       & 16112.64 & 76435.11 & .794                \\
                                                                                             &                & sphericity   & 1       & 1321.89  & 25083.02  & .240                \\
                                                                                             &                & \# clusters & 2       & 1089.54  & 10337.15  & .207                \\
                                                                                                                                            &                                  & \# vars     & 2       & 283.06   & 2685.57   & .063                \\
                                           & & \% categorical     & 2       & 282.99   & 2684.87   & .063                \\                                                                  &          & \# obs      & 2       & 185.60    & 1760.91    & .042                \\
                                            &                                & density     & 1       & 68.60  & 1301.74  & .016                \\
                                                         \cmidrule{2-7}
& \multirow{8}{*}{\begin{tabular}[c]{@{}l@{}}Within\\ data sets\\ effects\\ (univariate\\ tests)\end{tabular}} & Method (M)  & 7       & 4200.76  & 122234.65 & .606                \\
                                                                                             &                 & M*overlap   & 28       & 1113.81  & 8102.46  & .290                \\
   &                                       & M*categorical   & 14       & 413.61  & 6017.59  & .132                \\
                                                                     &      & M*clusters  & 14       & 328.71   & 4782.51  & .108                \\
                                                                                                                                                  &                 & M*obs       & 14       & 157.04    & 2284.78    & .054                \\
                               &    & M*sphericity   & 7       & 149.53    & 4350.96   & .052                \\
                                &               & M*vars      & 14       & 141.70   & 2061.73  & .049                \\
                                                                                                                                            &  & M*overlap*categorical       & 56       & 92.58   & 336.75  & .033    
                                                                                                                    \\
                                                                                         &  & M*overlap*density       & 28       & 82.16   & 597.71  & .029                               
                                                                                               \\  
                                                                                               &  & M*overlap*vars       & 56       & 73.77   & 268.31  & .026   
                                                                                               \\  
                                                                                                                            & & M*density       & 7       & 34.48   & 1555.65  & .019 
\\
\botrule
& Effect                                                                                                       & Source      & df      & SS      & $F$        & Partial $\eta^2$ \\ \midrule
\multirow{15}{*}{\begin{tabular}[c]{@{}l@{}}AMI\end{tabular}} & \multirow{7}{*}{\begin{tabular}[c]{@{}l@{}}Between\\ data sets\\ effects\end{tabular}}                       & overlap     & 4       & 15745.59 & 103229.18 & .839                \\
                                                                                             &                & sphericity  & 1       & 1217.15  & 31918.95  & .287                \\
                                                                                             &                & \# vars & 2       & 412.21  & 5405.00  & .120                \\
                                         &  & \% categorical     & 2       & 291.69   & 3824.67   & .088                \\                                                                       &     & \# obs      & 2       & 65.27    & 855.89    & .021                \\
                                           &                                 & density     & 1       & 61.67  & 1617.29  & .020                \\
                                                &                      & \# clusters & 2       & 9.20  & 120.58  & .003                \\      
                                                                \cmidrule{2-7}
& \multirow{8}{*}{\begin{tabular}[c]{@{}l@{}}Within\\ data sets\\ effects\\ (univariate\\ tests)\end{tabular}} & Method (M)  & 7       & 3778.28  & 170484.85 & .682                \\
                                                                                             &                & M*overlap   & 28       & 885.97  & 9994.29  & .335                \\
                         &                & M*categorical   & 14       & 345.71  & 7799.54  & .164                \\
                         &                                                  & M*clusters  & 14       & 220.68   & 4978.78  & .111               \\
                                                                                                                                            &                       & M*obs       & 14       & 145.00    & 3271.27    & .076                \\
                        &           & M*vars   & 14       & 116.96    & 2638.72   & .062                \\
                         &                      & M*sphericity      & 7       & 91.60   & 4133.20  & .049                \\
                                                                                                                                            &  & M*overlap*categorical       & 56       & 71.71  & 404.48  & .039    
                                                                                                                    \\
                                                                                         &  & M*overlap*vars       & 56       & 60.9   & 343.37  & .033                               
                                                                                               \\  
                                                                                               &  & M*overlap*density       & 28       & 45.48   & 513.09  & .025   
                                                                                               \\                 & & M*density       & 7       & 34.48   & 1555.65  & .019 
                                                               
\\
\botrule
\end{tabular}
\label{Tab:2}
\end{minipage}
\end{center}
\end{table}

The between data sets effects in Table \ref{Tab:2} can be thought of as the influence of the design factors across all clustering methods. Cluster overlap had the largest effect on cluster recovery. Overall, and as expected, as the overlap of clusters increased from $.01$ to $.20$, the average recovery in terms of both ARI/AMI decreased, going from $.55/.59$ to $.11/.16$. Cluster sphericity had also a large effect on cluster recovery with mean ARI/AMI values of $.32/.36$ for spherical and $.23/.27$ for non-spherical clusters. The number of clusters had a large and negative effect on cluster recovery based on ARI, with $.33$, $.27$ and $.22$, for $3$, $5$ and $8$ clusters, respectively; the effect was negligible in the case of AMI ($.32$, $.31$, $.32$). As the number of variables increased, the clustering performance deteriorated, as indicated by ARI/AMI values of $.30/.35$, $.27/.31$ and $.25/.29$ for $8$, $12$ and $16$ variables, respectively. The percentage of categorical variables in the data set had also a moderate and negative effect, with mean ARI/AMI values equal to $.29/.34$, $.28/.33$ and $.24/.29$, for $20\%$, $50\%$ and $80\%$, respectively. The number of observations had a more profound effect in the case of ARI, with mean values of $.25/.33$, $.28/.31$ and $.29/.31$ for $100$, $600$ and $1000$ observations, respectively. Last, cluster density had a small effect on cluster recovery, with mean ARI/AMI values equal to $.26/.31$ and $.28/.33$ for equally-sized clusters and a $10$\% of the observations in a single cluster, respectively. 

\begin{table}
\begin{center}
\begin{minipage}{\textwidth}
\caption{Cluster recovery (ARI/AMI) of eight clustering methods by cluster overlap, cluster sphericity, number of clusters, percentage of categorical variables, number of variables, cluster density and number of observations}
\centering
\rotatebox{90}{
\begin{tabular}{lrrrrrrrrr}
 \toprule
Factor                       & \multicolumn{1}{l}{Level} & \multicolumn{1}{l}{\begin{tabular}[c]{@{}l@{}}KAMILA\end{tabular}} & \multicolumn{1}{l}{\begin{tabular}[c]{@{}l@{}}FAMD \\ /KM\end{tabular}} & \multicolumn{1}{l}{\begin{tabular}[c]{@{}l@{}}K-Proto \\ types \end{tabular}} & \multicolumn{1}{l}{\begin{tabular}[c]{@{}l@{}}M-S \\ KM\end{tabular}} & \multicolumn{1}{l}{\begin{tabular}[c]{@{}l@{}}Mixed \\ RKM \end{tabular}} & \multicolumn{1}{l}{\begin{tabular}[c]{@{}l@{}}HL \\ /PAM\end{tabular}} &
\multicolumn{1}{l}{\begin{tabular}[c]{@{}l@{}}Mixed \\ KM\end{tabular}} &
\multicolumn{1}{l}{\begin{tabular}[c]{@{}l@{}}Gower \\ /PAM\end{tabular}} 
\\
 \midrule
\multirow{5}{*}{overlap} & .01 & .69/.72  & .65/.68  & .64/.67  & .64/.68 &  .64/.70 & .41/.47 & .39/.46 & .29/.34  \\
                          & .05 & .45/.49  & .42/.46  & .41/.46  & .39/.43 & .39/.44 &.23/.29  & .23/.29 & .16/.21 \\
                          & .10 & .31/.35 & .28/.33  & .28/.32  & .24/.29 & .24/.29 & .15/.20 & .15/.21 & .10/.15    \\
                          & .15 & .22/.26 & .20/.25  & .20/.24  & .17/.21 & .15/.20 & .10/.15 & .11/.16 & .07/.12  \\
                          & .20  & .16/.20  & .15/.19 & .15/.19 & .12/.17 & .10/.15 & .07/.12 & .08/.13  & .06/.10 \\
 \midrule
sphericity                    & no    & .30/.35  & .28/.33  & .29/.33   & .26/.31 & .17/.31  & .17/.22 & .17/.22 & .11/.15 \\
                             & yes   & .43/.46  & .40/.44  & .39/.43  & .37/.41  & .22/.41  & .22/.28 & .22/.27 & .17/.21    \\
\midrule
\multirow{3}{*}{\# clusters} & 3    & .42/.41   & .44/.42    & .39/.38   & .38/.37     & .40/.39     & .21/.22 & .21/.22 & .15/.16  \\
                             & 5    & .37/.40  & .34/.38     & .34/.38    & .30/.35     & .29/.34     & .19/.25 & .19/.25 & .14/.18 \\
                             & 8    & .31/.40   & .25/.35     & .28/.38    & .26/.36   & .24/.35     & .17/.28  & .17/.28 & .12/.21 \\
\midrule
\multirow{3}{*}{\% categorical} & 20\%  & .42/.45   & .35/.39     & .41/.44   & .30/.35   & .30/.35   & .21/.27 &  .24/.29 &.12/.17 \\
                             & 50\%   & .38/.41  & .34/.38     & .34/.38    & .35/.39   & .31/.36   & .19/.25  & .20/.26 & .13/.18 \\
                             & 80\%   & .30/.35   & .33/.37   & .26/.31  & .28/.32  & .32/.32   & .18/.36  & .14/.22 & .15/.20 \\
\midrule

\multirow{3}{*}{\# vars}     & 8    & .40/.44  & .34/.39  & .36/.41   & .33/.38    & .31/.37    & .25/.31  & .24/.30 & .17/.22\\
                             & 12  & .36/.40   & .34/.38 & .33/.37    & .30/.35     & .31/.36    & .19/.24 & .18/.24 & .13/.18\\
                             & 16  & .34/.37 & .33/.37   & .31/.35    & .30/.34     & .31/.35    & .14/.20 & .15/.20 & .11/.15\\
 \midrule
density                      & 10\%   & .35/.39   & .32/.37   & .31/.36   & .30/.34    & .29/.34    & .20/.25 & .19/.25 & .14/.18 \\
                             & equal  & .38/.42   & .36/.40   & .36/.39   & .33/.37    & .33/.37    & .19/.25 & .19/.25 & .14/.18\\
\midrule
\multirow{3}{*}{\# obs}      & 100   & .30/.37  & .29/.37  & .31/.39   & .29/.37  & .28/.37  & .20/.29 & .18/.27 & .14/.22  \\ 
                             & 600   & .39/.41  & .36/.38  & .35/.37   & .32/.35  & .32/.35  & .19/.23 & .19/.23 &  .14/.16 \\
                             & 1000  & .40/.42  & .37/.39  & .35/.37   & .32/.35  & .32/.35  & .19/.23 & .20/.23 &  .13/.16 \\
                              \botrule
\end{tabular}}
\label{Tab:3}
\end{minipage}
\end{center}
\end{table}

Based on the within data sets effects (Table \ref{Tab:2}) we determine which methods are effective under which conditions. We start by considering two-way interactions first. Then we discuss three-way interactions. In the presence of a significant interaction, main effects and lower order effects were ignored.

\begin{figure}[H]
  \centering
  \includegraphics[scale=0.7]{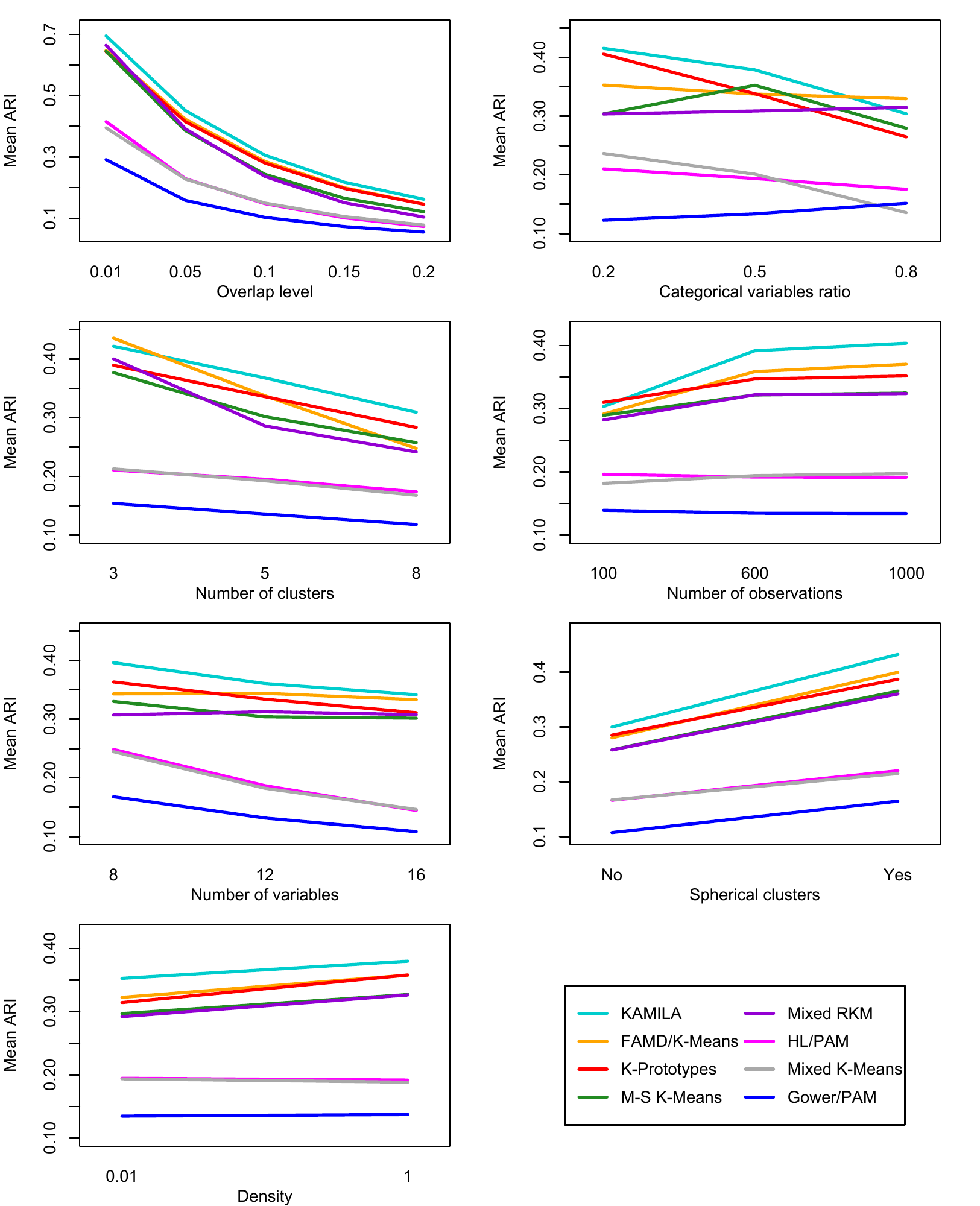}
  \caption{Two-way interactions of method by overlap level, percentage of categorical variables, number of clusters, number of observations, number of variables, density, and cluster sphericity (mean ARI values). Subplots/factors are arranged, from left to right, by decreasing effect size, partial $\eta^2$.}
  \label{Fig:3}
\end{figure}

Table \ref{Tab:3} shows the ARI/AMI values of the eight methods by all factors. In general, both measures yield similar results. The two-way interaction between method and number of clusters shows that the number of clusters negatively affects the performance of all methods, but the effect is more profound for Mixed Reduced K-Means, when the number of clusters is other than $3$ (Fig. \ref{Fig:3} and Table \ref{Tab:3}). The two-way interaction between method and number of observations reveals that K-Prototypes performs slightly better than other methods for the small sample-size scenario, $n = 100$ (Fig. \ref{Fig:3} and Table \ref{Tab:3}). KAMILA's performance greatly improves for $n = 600$. The performance of HL/PAM, Mixed K-Means and Gower/PAM does not appear to be significantly affected by $n$, but their performance remains poor compared to other methods. Non-sphericity of the clusters seems to affect all methods but not in a uniform manner (Fig. \ref{Fig:3} and Table \ref{Tab:3}). The difference in performance between KAMILA and other methods is less profound when clusters are non-spherical. This is not surprising, since in KAMILA continuous variables are assumed to follow a mixture distribution with arbitrary spherical clusters.

\begin{figure}[H]
  \centering
  \includegraphics[scale=0.55]{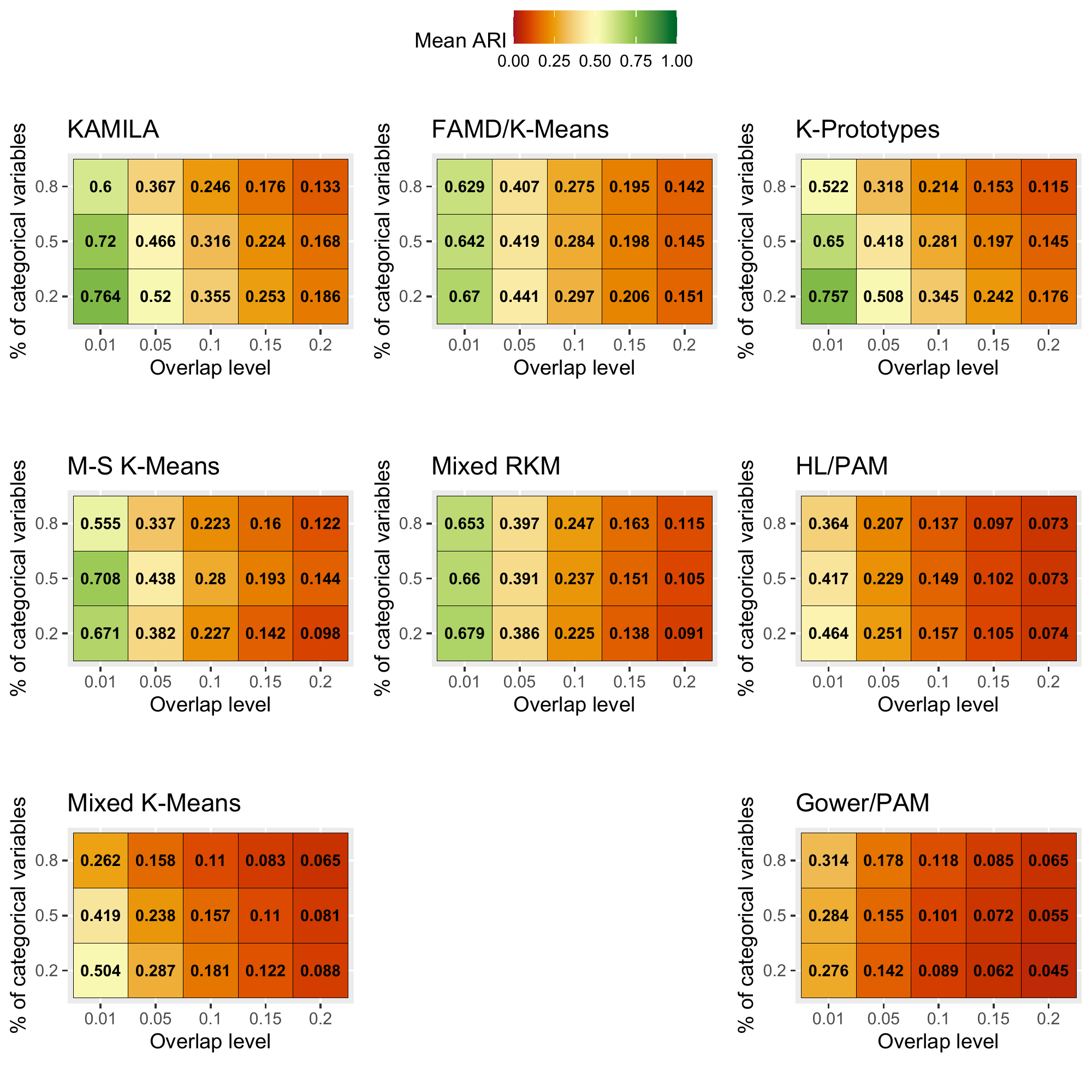}
  \caption{Three-way interaction of method by overlap level and percentage of categorical variables. The numbers indicate the mean ARI for each combination of overlap level and percentage of categorical variables.}
  \label{Fig:4}
\end{figure}

\begin{figure}[H]
  \centering
  \includegraphics[scale=0.55]{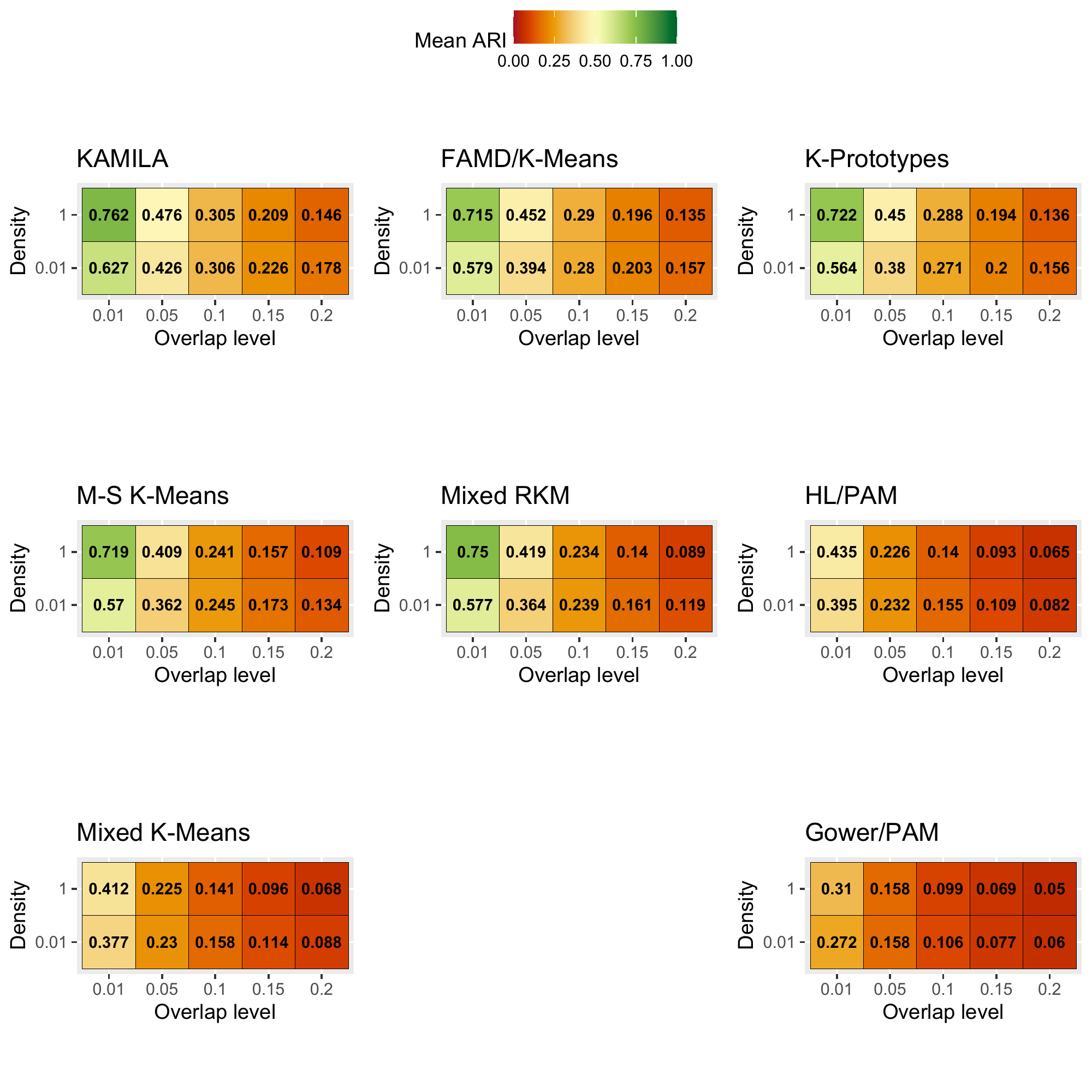}
  \caption{Three-way interaction of method by overlap level and cluster density. The numbers indicate the mean ARI for each combination of overlap level and cluster density.}
  \label{Fig:5}
\end{figure}

\begin{figure}[H]
  \centering
  \includegraphics[scale=0.55]{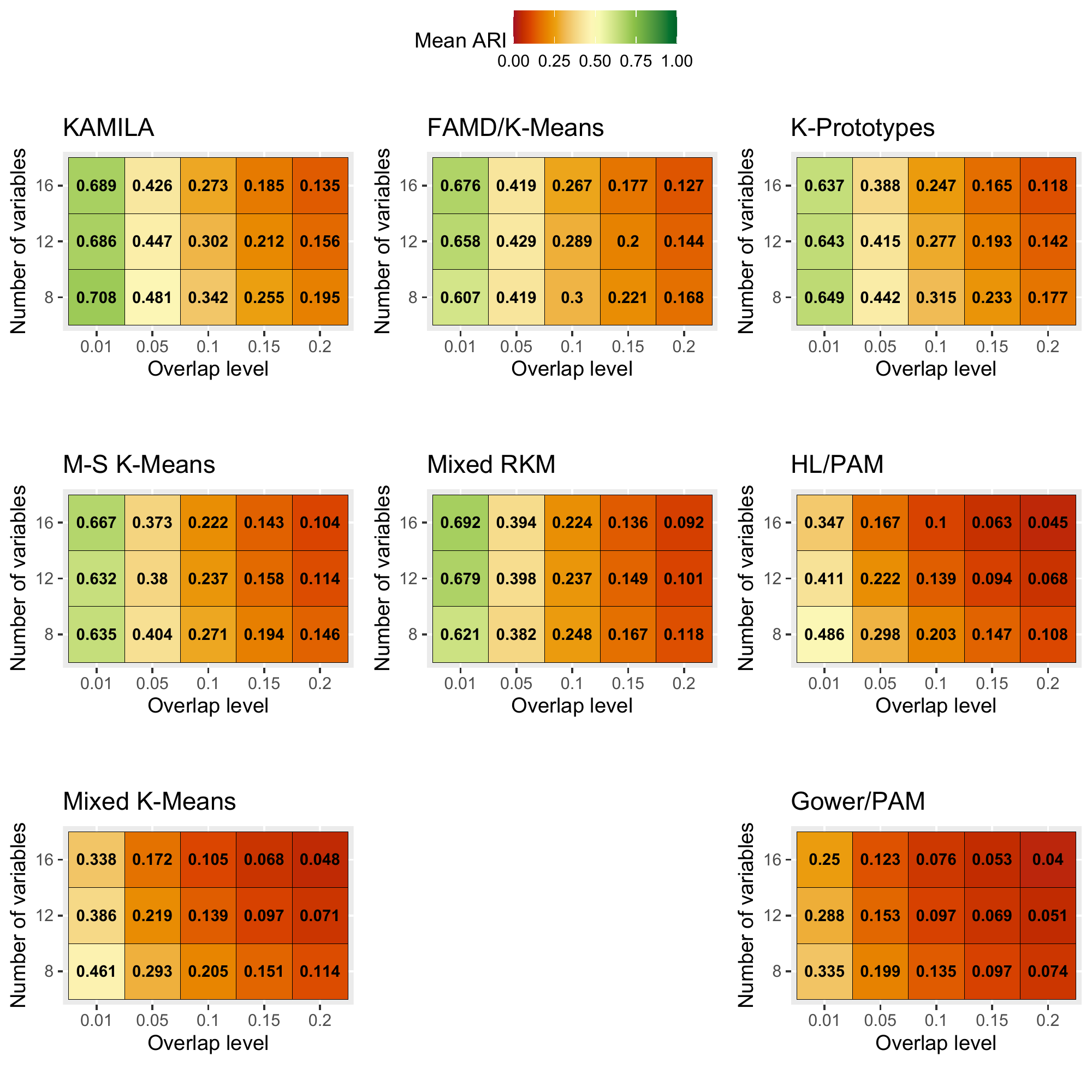}
  \caption{Three-way interaction of method by overlap level and number of variables. The numbers indicate the mean ARI for each combination of overlap level and number of variables.}
  \label{Fig:6}
\end{figure}

The heat maps in Figure \ref{Fig:4} visualize the three-way interaction of method by cluster overlap and percentage of categorical variables (mean ARI values). In the presence of categorical variables, there are differences between methods. KAMILA and K-Prototypes outperform other methods when the percentage of categorical variables is low ($20$\%), whereas FAMD/K-Means and Mixed Reduced K-Means perform best when the percentage of categorical variables is high ($80$\%). The performance of Mixed Reduced K-Means deteriorates at a faster rate than other methods with increasing overlap. Modha-Spangler K-Means performs best when the number of categorical and continuous variables in the data set is equal ($50$\%). The interaction of method by cluster overlap and density is illustrated in Figure \ref{Fig:5}. Some methods are affected more than others by the presence of a small-size cluster and at different levels of overlap. When cluster overlap is very low ($.01$) all methods perform better in the case of clusters with equal size, but for higher levels of overlap ($>.01$) there is negligible difference in cluster recovery or cluster recovery is slightly better in the small-size cluster scenario. The interaction of method by cluster overlap and the number of variables (Fig. \ref{Fig:6}) reveals that going from $8$ to $16$ variables, deteriorates the performance of KAMILA, K-Prototypes, HL/PAM, Mixed K-Means and Gower/PAM, whereas cluster recovery of Mixed Reduced K-Means and FAMD/K-Means is improved. This improvement, however, is observed only when cluster overlap is low ($\leq.05$).

Last, it is worth underlining that in terms of absolute performance, the mean ARI/AMI values in Table \ref{Tab:3} and Figures \ref{Fig:4} to \ref{Fig:3} suggest poor cluster recovery in the vast majority of cases (ARI/AMI values below $.65$) for all methods under comparison. Cluster recovery was found to be fair, albeit not good or excellent, for certain methods and conditions only (values above $.65$ but less than $.80$). In particular, for three out of eight methods (Mixed K-Means, HL/PAM, Gower/PAM) average cluster recovery is poor across all factors, even for well-separated and equally sized clusters. For the top-performing methods, there are cases when cluster recovery can be considered fair to good, especially when clusters are well-separated, equally sized and the percentage of categorical variables is low or moderate.

\section{Discussion}\label{Sec:6}

This paper reports benchmark test results from applying distance-based partitioning methods on simulated data sets with different characteristics. Eight methods were selected to cover three general strategies of distance or dissimilarity-based partitioning of mixed-type data (i.e., constructing a dissimilarity matrix between observations given as input to K-Medoids, extending K-Means to mixed-type data and reducing the number of variables and clustering of the observations in the reduced space).

One essential goal of the benchmark is to make the results available and reusable to other researchers. Benchmark results revealed both similarities and differences in the overall performance of the eight algorithms, as well as across different criteria. A group of top-performing methods with similar performance can be distinguished, consisting of KAMILA, FAMD/K-Means and K-Prototypes. KAMILA was the best method in about half of the $1620$ different data scenarios (based on both ARI and AMI). These are mostly data sets with moderate or large sample size and more continuous than categorical variables. The deterioration of KAMILA's performance in the small sample size scenario was somewhat expected, since the method's reliance on a multinomial model for categorical variables requires a commensurate sample size, as has been previously indicated in \cite{foss2019distance}. Therefore, we recommend the use of KAMILA when the sample size is reasonably large and the categorical variables are not dominant in the data set. FAMD/K-Means was the top-performing approach in about one fifth of the cases. This method performed well for data sets with moderate or large sample size and more categorical than continuous variables. In contrast to the other two methods, FAMD/K-Means additionally involves a dimensionality reduction step, which can be convenient for visualizing and interpreting the clusters in the reduced space. This means that it depends heavily on the amenability of the data set to dimensionality reduction where a few principal components account for a high percentage of variability in the data set.  Where this is not the case, the FAMD/K-Means method cannot be reasonably applied. K-Prototypes was the best approach in about $13$\% of the distinct scenarios and is recommended in cases when the sample size is relatively small and there are more continuous than categorical variables. 

Modha-Spangler K-Means and Mixed Reduced K-Means form a second group of methods with similar performance, not far from the first group in terms of cluster recovery. Modha-Spangler K-Means was the best method in $11\%$ of the different scenarios, mainly when the number of continuous variables is equal or greater than that of categorical variables. This is also demonstrated in \cite{foss2016semiparametric}, where Modha-Spangler K-Means was found to underperform relative to competing methods when there are more categorical than continuous variables because of its over-reliance on continuous variables. Mixed Reduced K-Means was the best approach in $7\%$ of the cases, performing well for moderate or large samples sizes, more categorical than continuous variables and low levels of cluster overlap. Although Mixed Reduced K-Means was expected to improve upon FAMD/K-Means, in the sense that it was developed to address the cluster masking problem by optimizing a single objective function \citep{vichi2019clustering}, this hypothesis was not confirmed by the study results. Therefore, when dimensionality reduction is an additional goal, FAMD/K-Means seems to be a more reasonable choice than Mixed Reduced K-Means. However, both FAMD/K-Means and Mixed Reduced K-Means have in common that they perform worse for a sample size of $100$ compared to larger sample sizes; a sample size of $100$ is usually not sufficient for PCA-based methods \cite[see e.g.,][]{saccenti2016approaches}. 

A third group of methods, clearly distinct from the other two in terms of cluster recovery, contains HL/PAM, Mixed K-Means and Gower/PAM. These methods demonstrated poor performance in our experiments, with Gower/PAM being the worst performing method across all criteria, even for well-separated clusters. This could be seen as a surprising finding, considering that Gower/PAM is among the most popular choices in the literature for clustering mixed-type data. A potential explanation for this could be that the task of clustering multivariate normal distributions, as the objective of the simulations conducted in the current study, can be expected to favour K-Means-like approaches that use a squared loss function; PAM-based approaches instead are known to be more robust against non-normality \citep[p.117]{kaufman2009finding}. Also notice that HL/PAM performed better than Gower/PAM in all tested scenarios but did not reach the performance of K-Means-based methods.

There are some limitations with the current study. First, to generate mixed-type data for the simulations, continuous variables were generated by drawing from finite mixtures of multivariate normal distributions; categorical variables were generated via discretization of such continuous variables. Ideally, for our experiments we would need to generate purely mixed-type data, that is, purely categorical variables and purely continuous variables with a cluster structure. However, controlling the overlap between clusters in mixed-type data sets with more than two clusters is not straightforward \citep[see, e.g.,][]{maitra2010simulating}. In addition, the covariance structure between the variables was not user-defined. Controlling the correlation structure between the variables could have been useful, so as to draw conclusions on how correlated variables affect the performance of clustering algorithms. Also, the simulations could be extended to mixtures of non-Gaussian distributions. Second, clustering performance depends on the software implementation used; different implementations of a method often lead to different results. The included clustering methods were required to have an R-implementation that can be used in a default way without additional tuning in order to allow for a comparison that is not influenced by different tuning flexibilities. Furthermore, in a study by \cite{steinley2006profiling}, the K-Means algorithm with 100 random initializations has been shown to produce the same solution for well-separated clusters, but the algorithm produced different solutions in the case of overlapping clusters. The author recommended using several thousand random initializations. Based on this observation, the $100$ random starts used in our study might not be sufficiently high for K-Means-based clustering methods to avoid local minima. However, from a computational viewpoint, this would result in a much higher and prohibitive computational cost. Another limitation is that for the methods under comparison, the number of clusters was always correctly specified, instead of incorporating cluster number estimation in the clustering task. Although this is a challenging and interesting endeavour, it requires additional decisions, such as the choice of the index/method to be used for cluster estimation and was beyond the scope of the current study. Last, the results of the current study were based on simulated data sets only. An empirical comparison of cluster analysis methods on real mixed-type data \citep[see, e.g.,][]{hennig2022} is expected to further highlight how different clustering methods produce solutions with different data analytic characteristics, which can help a user choosing an appropriate method for the research question of interest.

\bibliography{main.bib}

\begin{thebibliography}{49}
\providecommand{\natexlab}[1]{#1}
\providecommand{\url}[1]{{#1}}
\providecommand{\urlprefix}{URL }
\providecommand{\doi}[1]{\url{https://doi.org/#1}}
\providecommand{\eprint}[2][]{\url{#2}}
 \bibcommenthead

\bibitem[{Ahmad and Dey(2007)}]{ahmad2007k}
Ahmad A, Dey L (2007) A k-mean clustering algorithm for mixed numeric and
  categorical data. Data \& Knowledge Engineering 63(2):503--527

\bibitem[{Ahmad and Khan(2019)}]{ahmad2019survey}
Ahmad A, Khan SS (2019) Survey of state-of-the-art mixed data clustering
  algorithms. IEEE Access 7:31,883--31,902

\bibitem[{Arabie(1994)}]{arabie1994cluster}
Arabie P (1994) Cluster analysis in marketing research, Blackwell, Oxford, pp
  160--189

\bibitem[{Audigier et~al.(2016)Audigier, Husson, and
  Josse}]{audigier2016principal}
Audigier V, Husson F, Josse J (2016) A principal component method to impute
  missing values for mixed data. Advances in Data Analysis and Classification
  10(1):5--26

\bibitem[{Boulesteix and Hatz(2017)}]{boulesteix2017}
Boulesteix AL, Hatz M (2017) Benchmarking for clustering methods based on real
  data: A statistical view. In: Palumbo F, Montanari A, Vichi M (eds) Data
  Science. Springer International Publishing, Cham, pp 73--82

\bibitem[{Boulesteix et~al.(2013)Boulesteix, Lauer, and
  Eugster}]{boulesteix2013}
Boulesteix AL, Lauer S, Eugster MJ (2013) A plea for neutral comparison studies
  in computational sciences. PLoS ONE 8:e61,562

\bibitem[{De~Soete and Carroll(1994)}]{desoete1994k}
De~Soete G, Carroll JD (1994) K-means clustering in a low-dimensional Euclidean
  space, Springer, pp 212--219

\bibitem[{Dolnicar and Gr{\"u}n(2008)}]{dolnicar2008challenging}
Dolnicar S, Gr{\"u}n B (2008) Challenging ``factor--cluster segmentation''.
  Journal of Travel Research 47(1):63--71

\bibitem[{Ferreira and Hitchcock(2009)}]{ferreira2009}
Ferreira L, Hitchcock DB (2009) A comparison of hierarchical methods for
  clustering functional data. Communications in Statistics - Simulation and
  Computation 38(9):1925--1949

\bibitem[{Foss et~al.(2016)Foss, Markatou, Ray, and
  Heching}]{foss2016semiparametric}
Foss A, Markatou M, Ray B, et~al. (2016) A semiparametric method for clustering
  mixed data. Machine Learning 105(3):419--458

\bibitem[{Foss and Markatou(2018)}]{foss2018kamila}
Foss AH, Markatou M (2018) kamila: Clustering mixed-type data in {R} and
  {H}adoop. Journal of Statistical Software 83:1--44

\bibitem[{Foss et~al.(2019)Foss, Markatou, and Ray}]{foss2019distance}
Foss AH, Markatou M, Ray B (2019) Distance metrics and clustering methods for
  mixed-type data. International Statistical Review 87(1):80--109

\bibitem[{Gower(1971)}]{gower1971general}
Gower JC (1971) A general coefficient of similarity and some of its properties.
  Biometrics 27:857--871

\bibitem[{Hennig(2020)}]{hennig2015packagefpc}
Hennig C (2020) Package `fpc'.
  \urlprefix\url{https://https://cran.r-project.org/web/packages/fpc/fpc.pdf}

\bibitem[{Hennig(2022)}]{hennig2022}
Hennig C (2022) An empirical comparison and characterisation of nine popular
  clustering methods. Advances in Data Analysis and Classification 16:201--229

\bibitem[{Hennig and Liao(2013)}]{hennig2013find}
Hennig C, Liao TF (2013) How to find an appropriate clustering for mixed-type
  variables with application to socio-economic stratification. Journal of the
  Royal Statistical Society: Series C (Applied Statistics) 62(3):309--369

\bibitem[{Huang(1997)}]{huang1997clustering}
Huang Z (1997) Clustering large data sets with mixed numeric and categorical
  values. In: Proceedings of the 1st {P}acific-{A}sia Conference on Knowledge
  Discovery and Data Mining (PAKDD), Citeseer, pp 21--34

\bibitem[{Hubert and Arabie(1985)}]{hubert1985}
Hubert L, Arabie P (1985) Comparing partitions. Journal of Classification
  2(2):193--218

\bibitem[{Hunt and Jorgensen(2011)}]{hunt2011clustering}
Hunt L, Jorgensen M (2011) Clustering mixed data. Wiley Interdisciplinary
  Reviews: Data Mining and Knowledge Discovery 1(4):352--361

\bibitem[{Javed et~al.(2020)Javed, Lee, and Rizzo}]{javed2020}
Javed A, Lee BS, Rizzo DM (2020) A benchmark study on time series clustering.
  Machine Learning with Applications 1:100,001

\bibitem[{Jimeno et~al.(2021)Jimeno, Roy, and Tortora}]{jimeno2021}
Jimeno J, Roy M, Tortora C (2021) Clustering mixed-type data: A benchmark study
  on {KAMILA} and {K-P}rototypes. In: Chadjipadelis T, Lausen B, Markos A,
  et~al. (eds) Data Analysis and Rationality in a Complex World. Springer
  International Publishing, Cham, pp 83--91

\bibitem[{Kaufman and Rousseeuw(1990)}]{kaufman2009finding}
Kaufman L, Rousseeuw PJ (1990) Finding {G}roups in {D}ata: {A}n {I}ntroduction
  to {C}luster {A}nalysis, John Wiley \& Sons, chap~2, pp 68--125

\bibitem[{Kiers(1991)}]{kiers1991simple}
Kiers HA (1991) Simple structure in component analysis techniques for mixtures
  of qualitative and quantitative variables. Psychometrika 56(2):197--212

\bibitem[{Maechler et~al.(2021)Maechler, Rousseeuw, Struyf, Hubert, and
  Hornik}]{Maechler2021}
Maechler M, Rousseeuw P, Struyf A, et~al. (2021) cluster: Cluster Analysis
  Basics and Extensions.
  \urlprefix\url{https://CRAN.R-project.org/package=cluster}, {R} package
  version 2.1.2)

\bibitem[{Maitra and Melnykov(2010)}]{maitra2010simulating}
Maitra R, Melnykov V (2010) Simulating data to study performance of finite
  mixture modeling and clustering algorithms. Journal of Computational and
  Graphical Statistics 19(2):354--376

\bibitem[{Markos et~al.(2019)Markos, Iodice~D'Enza, and van~de
  Velden}]{markos2019beyond}
Markos A, Iodice~D'Enza A, van~de Velden M (2019) Beyond tandem analysis: Joint
  dimension reduction and clustering in {R}. Journal of Statistical Software
  91:1--24

\bibitem[{Markos et~al.(2020)Markos, Moschidis, and
  Chadjipantelis}]{markos2020sequential}
Markos A, Moschidis O, Chadjipantelis T (2020) Sequential dimension reduction
  and clustering of mixed-type data. International Journal of Data Analysis
  Techniques and Strategies 12(3):228--246

\bibitem[{Meilă and Heckerman(2001)}]{meil2001}
Meilă M, Heckerman D (2001) An experimental comparison of model-based
  clustering methods. Machine Learning 42:9--29

\bibitem[{Melnykov and Maitra(2010)}]{melnykov2010}
Melnykov V, Maitra R (2010) Finite mixture models and model-based clustering.
  Statistics Surveys 4:80--116

\bibitem[{Melnykov et~al.(2012)Melnykov, Chen, and Maitra}]{melnykov2012}
Melnykov V, Chen WC, Maitra R (2012) Mix{S}im: An \uppercase{R} package for
  simulating data to study performance of clustering algorithms. Journal of
  Statistical Software 51(12):1--25

\bibitem[{Milligan(1980)}]{milligan1980}
Milligan GW (1980) An examination of the effect of six types of error
  perturbation on fifteen clustering algorithms. Psychometrika 45:325--342

\bibitem[{Modha and Spangler(2003)}]{modha2003feature}
Modha DS, Spangler WS (2003) Feature weighting in k-means clustering. Machine
  Learning 52(3):217--237

\bibitem[{Murtagh(2015)}]{murtagh2015brief}
Murtagh F (2015) A {B}rief {H}istory of {C}luster {A}nalysis. In: Hennig C,
  Meila M, Murtagh F, et~al. (eds) Handbook of Cluster Analysis. Chapman \&
  Hall/CRC, pp 21--33

\bibitem[{Pag{\`e}s(2014)}]{pages2014multiple}
Pag{\`e}s J (2014) Multiple Factor Analysis By Example Using R, Chapman and
  Hall/CRC, chap~3, pp 67--78

\bibitem[{Preud’Homme et~al.(2021)Preud’Homme, Duarte, Dalleau, Lacomblez,
  Bresso, Sma{\"\i}l-Tabbone, Couceiro, Devignes, Kobayashi, Huttin
  et~al.}]{preud2021}
Preud’Homme G, Duarte K, Dalleau K, et~al. (2021) Head-to-head comparison of
  clustering methods for heterogeneous data: a simulation-driven benchmark.
  Scientific Reports 11(1):1--14

\bibitem[{Saccenti and Timmerman(2016)}]{saccenti2016approaches}
Saccenti E, Timmerman ME (2016) Approaches to sample size determination for
  multivariate data: Applications to {PCA} and {PLS-DA} of omics data. Journal
  of Proteome Research 15(8):2379--2393

\bibitem[{Saraçli et~al.(2013)Saraçli, Doğan, and İsmet
  Doğan}]{sarali2013}
Saraçli S, Doğan N, İsmet Doğan (2013) Comparison of hierarchical cluster
  analysis methods by cophenetic correlation. Journal of Inequalities And
  Applications 2013:1--8

\bibitem[{Shireman et~al.(2016)Shireman, Steinley, and
  Brusco}]{shireman2016local}
Shireman EM, Steinley D, Brusco MJ (2016) Local optima in mixture modeling.
  Multivariate Behavioral Research 51(4):466--481

\bibitem[{Steinley(2004)}]{steinley2004properties}
Steinley D (2004) Properties of the {H}ubert-{A}rabie {A}djusted {R}and
  {I}ndex. Psychological Methods 9(3):386--396

\bibitem[{Steinley(2006)}]{steinley2006profiling}
Steinley D (2006) Profiling local optima in k-means clustering: developing a
  diagnostic technique. Psychological Methods 11(2):178--192

\bibitem[{Szepannek(2018)}]{Szepannek2018}
Szepannek G (2018) clust{M}ix{T}ype: User-{F}riendly {C}lustering of
  {M}ixed-{T}ype {D}ata in {R}. The R Journal 10(2):200--208

\bibitem[{Van~{der Hoef} and Warrens(2019)}]{vanderhoef2019}
Van~{der Hoef} H, Warrens MJ (2019) Understanding information theoretic
  measures for comparing clusterings. Behaviormetrika 46:353--370

\bibitem[{Van~Mechelen et~al.(2018)Van~Mechelen, Boulesteix, Dang, Dean, Guyon,
  Hennig, Leisch, and Steinley}]{whitepaper}
Van~Mechelen I, Boulesteix AL, Dang R, et~al. (2018) Benchmarking in cluster
  analysis: A white paper \urlprefix\url{https://arxiv.org/abs/1809.10496v2}

\bibitem[{van~de Velden et~al.(2017)van~de Velden, Iodice~D'Enza, and
  Palumbo}]{van2017cluster}
van~de Velden M, Iodice~D'Enza A, Palumbo F (2017) Cluster correspondence
  analysis. Psychometrika 82(1):158--185

\bibitem[{van~de Velden et~al.(2019)van~de Velden, Iodice~D'Enza, and
  Markos}]{van2019distance}
van~de Velden M, Iodice~D'Enza A, Markos A (2019) Distance-based clustering of
  mixed data. Wiley Interdisciplinary Reviews: Computational Statistics
  11(3):e1456

\bibitem[{Vichi and Kiers(2001)}]{vichi2001factorial}
Vichi M, Kiers HA (2001) Factorial k-means analysis for two-way data.
  Computational Statistics \& Data Analysis 37(1):49--64

\bibitem[{Vichi et~al.(2019)Vichi, Vicari, and Kiers}]{vichi2019clustering}
Vichi M, Vicari D, Kiers HA (2019) Clustering and dimension reduction for mixed
  variables. Behaviormetrika 46(2):243--269

\bibitem[{Vinh et~al.(2010)Vinh, Epps, and Bailey}]{vinh2010information}
Vinh NX, Epps J, Bailey J (2010) Information theoretic measures for clusterings
  comparison: Variants, properties, normalization and correction for chance.
  The Journal of Machine Learning Research 11:2837--2854

\bibitem[{Yamamoto and Hwang(2014)}]{yamamoto2014general}
Yamamoto M, Hwang H (2014) A general formulation of cluster analysis with
  dimension reduction and subspace separation. Behaviormetrika 41(1):115--129

\end{thebibliography}


\end{document}